\documentclass[journal=jacsat,dvipsnames,manuscript=article]{achemso}

\usepackage[T1]{fontenc} 
\usepackage{amsmath,amsfonts,amssymb, mathtools}
\usepackage{geometry}
\usepackage{mathrsfs}
\usepackage{physics}
\usepackage{setspace}
\usepackage{bbold}
\usepackage{calc}
\usepackage{subcaption}
\usepackage{array}
\usepackage{textcomp}
\usepackage{gensymb}
\usepackage{subfiles} 
\usepackage{lipsum}  
\usepackage{hyperref} 
\usepackage{siunitx}
\usepackage{booktabs}
\usepackage{todonotes}
\usepackage{caption}

\usepackage{color,soul}
\usepackage{xcolor}
\usepackage{comment}
\usepackage{algorithm}
\usepackage{algpseudocode}

\usepackage{siunitx}
\usepackage{xr}

\usepackage{hyperref}
\usepackage{xr-hyper}
\usepackage[version=4]{mhchem}

\author{Matteo Cartiglia}
\email{matteopero.cartiglia@imec.be}
\altaffiliation{These authors contributed equally}
\affiliation[imec]{imec, Kapeldreef 75, 3001 Heverlee, Belgium}
\author{Sandro Kuppel}
\altaffiliation{These authors contributed equally}
\affiliation[imec]{imec, Kapeldreef 75, 3001 Heverlee, Belgium}
\author{Wouter Botermans}
\altaffiliation{These authors contributed equally}
\affiliation[imec]{imec, Kapeldreef 75, 3001 Heverlee, Belgium}
\author{Wannes Peeters}
\affiliation[imec]{imec, Kapeldreef 75, 3001 Heverlee, Belgium}
\author{Natan Biesmans}
\affiliation[imec]{imec, Kapeldreef 75, 3001 Heverlee, Belgium}
\author{Liam Vandekerckhove}
\affiliation[imec]{imec, Kapeldreef 75, 3001 Heverlee, Belgium}
\author{Eric Beamish}
\affiliation[imec]{imec, Kapeldreef 75, 3001 Heverlee, Belgium}
\author{Koen Ongena}
\affiliation[imec]{imec, Kapeldreef 75, 3001 Heverlee, Belgium}
\author{Wouter Renckens}
\affiliation[imec]{imec, Kapeldreef 75, 3001 Heverlee, Belgium}
\author{Pol Van Dorpe}
\affiliation[imec]{imec, Kapeldreef 75, 3001 Heverlee, Belgium}
\alsoaffiliation{Department of Physics and Astronomy, KU Leuven, Celestijnenlaan 200D, B-3001 Leuven, Belgium}
\author{Sanjin Marion}
\email{sanjin.marion@imec.be}
\affiliation[imec]{imec, Kapeldreef 75, 3001 Heverlee, Belgium}

\title[]{Latent space mapping of interpretable structural coordinates from stochastic single-molecule signals}

\begin{document}

\begin{abstract}

Nanopores are versatile single-molecular sensors, but their utility is fundamentally constrained by stochastic translocation dynamics warping any encoded information.
We resolve it by shifting from time-domain analysis to a learned latent-space mapping via a contrastive encoder trained exclusively on simulated signals from a physics-informed model. This encoder maps solid-state nanopore signals of engineered DNA barcodes into an interpretable molecular coordinate system. 
The learned representation is responsive to structural barcode parameters while remaining invariant to acquisition conditions and translocation conformation, allowing data pooling across devices.
Molecule identification requires a single pass through the encoder, reducing computational cost by three orders of magnitude relative to alignment-based methods. We experimentally validate through mixture quantification, rare-variant detection, consensus barcode reconstruction, and real-time signal acquisition.
This shift from temporal analysis to mapping structural coordinates into a latent space changes the paradigm behind analyzing stochastic sensor signals by linking classification to interpretable encoded molecular information.

\end{abstract}

\section{Main}
The translocation of charged polymers through nanometer-scale apertures is a canonical problem in non-equilibrium statistical physics and the foundational mechanism of single-molecule nanopore sensing~\cite{muthukumar2011polymer, dekker2007solid, xue2020solid, butler2008single}. 
Solid-state nanopores (SSNPs), fabricated from robust materials such as silicon nitride ($\text{SiN}_x$), offer a tunable, abiotic platform for structural characterization of intact macromolecules across a broad range of experimental conditions~\cite{storm2005fast, wanunu2010electrostatic}. 
When analyzing engineered dsDNA barcodes, where structural labels such as dumbbells or single-stranded overhangs locally increase volume and charge~\cite{bell2016digitally, beamish2017identifying, boskovic2022nanopore}, each label produces a characteristic transient reduction in the ionic current as the molecule passes through the pore. The ionic-current time series recorded during one translocation event, the translocation trace (Fig.~\ref{fig:fig1}A), exhibits a pattern of dips that encodes the spatial arrangement of labels along the molecule, providing an electrically transduced readout of molecular structure that is amenable to integration and parallelization~\cite{koch2023nanopore, doroschak2020porcupine}.
Fast and robust recovery of such encoded information is key to enable nanopore sensing in fields such as DNA data storage readout~\cite{Doricchi2022}, biomarker detection~\cite{xue2020solid,bell2016digitally,bovskovic2023simultaneous}, mapping positions of sequence-specific bound proteins~\cite{yang2018detection, bulushev2016single}, and direct sequence fingerprinting of proteins~\cite{soni2025protein}.

\subsection{Temporal Warping Distorts the Single-Molecule Signal}

\begin{figure}
\centering
\includegraphics[width=\linewidth]{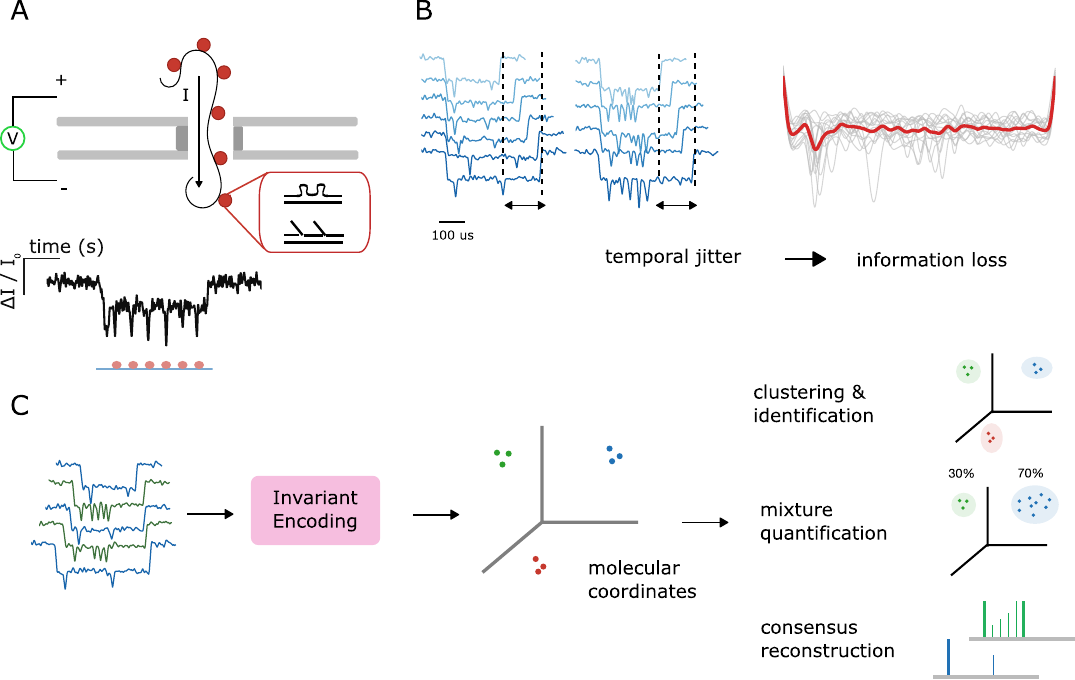}
\caption{\small \textbf{Nanopore sensing of DNA barcodes and the Latent Space Mapping (LSM) framework.}
(A)Schematic of solid-state nanopore sensing. A voltage bias drives barcoded (labeled) DNA through the pore. Structural labels, dumbbell hairpins or single-stranded polyT overhangs (red inset), locally increase volume and charge as they pass, producing transient dips in the ionic-current trace (bottom) that encode the spatial arrangement of labels along the molecule.
(B) The stochastic temporal warping problem. Left: Overlaid translocations of the same molecule highlight severe temporal fluctuations caused by nanopore kinetics, producing non-linearly time-distorted signal copies. Right: Superimposing 15 individual events (grey) demonstrates that naive temporal interpolation and averaging (red) destroys the underlying structural barcode, necessitating a velocity-invariant approach.
(C) Overview of our LSM framework. An encoder maps warped traces into a velocity-invariant latent space of molecular coordinates, clustering molecules by structural identity regardless of acquisition conditions. This new representation enables clustering and identification, quantification of mixtures, and barcode reconstruction.
}
\label{fig:fig1}
\end{figure}

Despite their versatility, solid-state nanopores face a fundamental physical challenge: the translocation process is inherently stochastic. The threading and translocation of the dsDNA polymer through a nanoscale constriction is governed by an interplay of different forces complicated by the evolution of the conformational states of the polymer coil~\cite{muthukumar2011polymer, storm2005fast, saito2011dynamical, chen2021dynamics, sakaue2007nonequilibrium}. Electrophoretic driving forces, hydrodynamic drag from electroosmosis, tension propagation, nanoscale friction, and molecular motion produce a complex and inherently stochastic translocation dynamic. This stochasticity manifests as temporal warping: the molecule's translocation velocity profile fluctuates widely, not only between distinct molecules but also between successive translocations of the same molecule~\cite{Plesa2015velocity}. Additionally, solid-state nanopores have inherent manufacturing variability in their sensing region, further complicating the pooling of data from nanopore arrays~\cite{xue2020solid}. The polymer may also enter the pore in a folded conformation, further distorting the measured current signature. 
Consequently, the ionic current signal of a labeled molecule (Fig.~\ref{fig:fig1}A) becomes a non-linearly distorted representation of the analyte's spatial configuration, in which the inter-label spacings that define the barcode are warped non-uniformly from event to event.

Na\"ive temporal interpolation and averaging of traces loses the structural features of the
underlying barcode (Fig. 1B); each translocation is a non-equilibrium event, and averaging
need not converge to a representative signal. 
This can be reduced by translocation velocity control, e.g., by enzymatic ratcheting in biological nanopores~\cite{manrao2012reading} or direct manipulation of single molecules~\cite {Leitao2023}, albeit at the cost of compromising molecular throughput. 
Ratcheting, however, has no counterpart in solid-state nanopores, which lack a motor protein to regulate transport; solid-state experimental approaches only partially mitigate stochasticity by tuning conditions at the pore~\cite{wanunu2010electrostatic, Kowalczyk2012, chou2024coupled} to decelerate translocation, without removing the underlying velocity fluctuations.

\subsection{A Coordinate System for Molecular Structure}

Traditional approaches to molecular identification, such as Dynamic Time Warping (DTW) and Hidden Markov Models (HMM), align signals in the time domain~\cite{schreiber2015analysis}, but degrade in highly stochastic free-translocation regimes and scale poorly to high-bandwidth experiments~\cite{misiunas2018quipunet, bao2021squigglenet, cao2024deep, dematties2021deep}.

Here we introduce Latent Space Mapping (LSM), a framework that replaces time-domain alignment with a learned molecular coordinate system (Fig.~\ref{fig:fig1}C). 
A Siamese ResNet encoder~\cite{koch2015siamese, khosla2020supervised}, trained exclusively on simulated signals from a physics-informed forward model, maps translocation traces into an embedding space where traces sharing the same barcode structure converge regardless of translocation velocity.
Contrastive synthetic-to-real transfer has proven effective in computer vision~\cite{tobin2017domain, chen2021contrastive}, but has not been explored for physically stochastic sensor signals.
This representation enables downstream tasks including clustering, mixture quantification from cluster populations, rare-variant detection at abundances as low as 0.2\%, and consensus barcode reconstruction via a Transformer-based Multi-Instance Learner (MIL)~\cite{dietterich1997solving, ilse2018attention, vaswani2017attention, zaheer2017deep, lee2019set}. Identification requires only a single forward pass through the encoder, enabling these analyses to scale to real-time deployment on multichannel readers.


\begin{figure}

\centering
\includegraphics[width=0.75\linewidth]{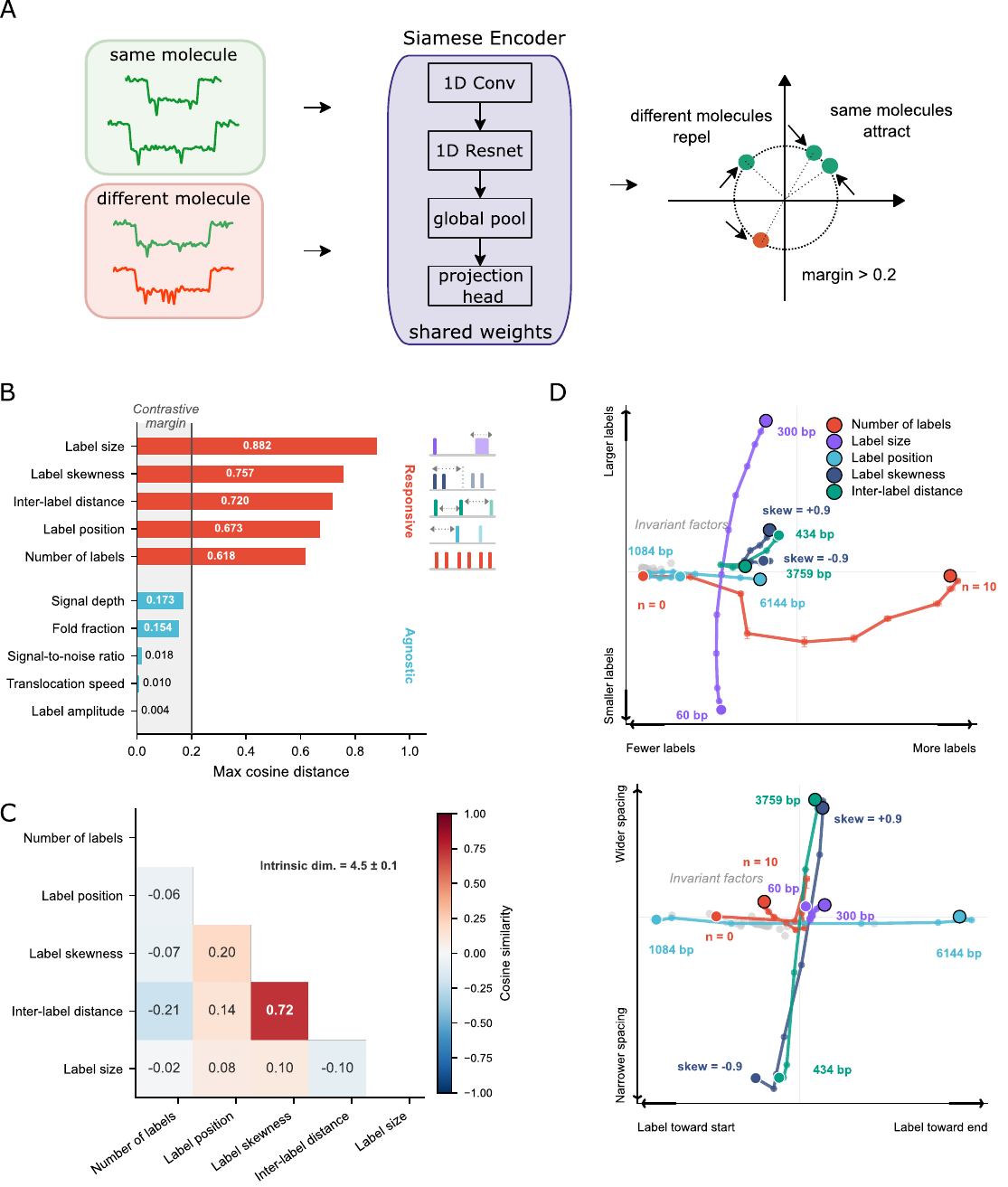}
\caption{\small \textbf{Contrastive encoder architecture and latent space organization.}
(A) The Siamese encoder maps pairs of translocation traces through a 1D ResNet into a 256-dimensional embedding space. A contrastive cosine similarity loss minimizes the cosine distance between embeddings of traces from the same molecule while maximizing it between traces from different molecules, enforcing a minimum separation margin of 0.2.
(B) Encoder responsiveness to barcode design factors versus acquisition and conformation factors. Each bar reports the maximum pairwise cosine distance between any two per-level mean embeddings when sweeping a single factor across its full range while the remaining factors are sampled from their default distributions. 
(C) Disentanglement of the five structural axes. For each structural factor, the trajectory of mean embeddings across the OFAT sweep defines a direction in latent space (its first principal component). The matrix shows the pairwise cosine similarities between these five directions. The skewness–inter-label distance pair has a cosine similarity of 0.72; all other pairs are $\leq 0.21$. An independent TwoNN estimate of the intrinsic dimensionality is $4.5 \pm 0.1$, 95\% CI.
(D) OFAT sweeps projected onto two-dimensional slices of a four-dimensional functional basis. Each axis is the dominant direction of one structural factor's sweep through embedding space, computed as the first principal component of its per-level mean embeddings; the four axes are orthogonalized with the Gram–Schmidt orthogonalization. Each colored trajectory shows the mean embedding ($\pm1 s.e.m.$ n = 50 traces per level) as one barcode parameter is swept across its range. Gray points show the corresponding acquisition and conformation factor sweeps.}
\label{fig:fig2}
\end{figure}

\subsection{Anatomy of the Coordinate System}

The Siamese ResNet encoder maps each raw translocation trace to a fixed point in a 256-dimensional latent space (Fig.~\ref{fig:fig2}A). A contrastive cosine similarity loss enforces a minimum separation margin of $m = 0.2$ between embeddings of traces from different molecules (see Methods) and minimizes the distance between traces of the same molecule.
The encoder is trained on signals from a physics-informed forward model that simulates both linear and folded translocation events, so that both conformations map to the same latent region (Methods and Supplementary Section~\ref{sec:sm_synthetic_detail}).

Because the forward model exposes each generative factor as an independently controllable parameter, the learned latent space can be interrogated directly. Ten factors split into two groups: five structural (label count, size, position, skewness, and inter-label distance) and five non-structural (signal depth, SNR, translocation speed, label amplitude, and fold fraction).

To test which generative factors the encoder is responsive to, and whether that responsiveness separates structural from acquisition parameters, we performed a one-factor-at-a-time (OFAT) analysis: varying each factor while sampling the others from their default distributions, and measuring the resulting displacement in the embedding (see Methods and Supplementary Section~\ref{sec:sm_ofat}).

As shown in Fig.~\ref{fig:fig2}B, the five structural factors produce embedding shifts well above the contrastive margin, while all non-structural factors remain below it (Supplementary Fig.~\ref{fig:sm_ofat_responsiveness}). The encoder correctly distinguishes physically similar quantities: label size (longitudinal extent of a blockade, intrinsic to the dumbbell) is treated as structural, whereas label amplitude (blockade depth, dependent on pore geometry and bias) is treated as an acquisition nuisance and left invariant.

To assess whether these five structural factors are encoded independently, we computed the pairwise cosine similarity between directions each factor traces through embedding space during its OFAT sweep (Fig.~\ref{fig:fig2}C). Most pairs are nearly orthogonal ($\leq 0.21$), confirming that the encoder represents each structural property in an independent subspace. The one exception is the skewness--inter--label distance pair (cosine similarity $\approx 0.72$), whose coupling reflects their shared geometric content: both factors describe the relative spatial arrangement of labels on the scaffold. The five structural factors, with one partially coupled pair, therefore yield approximately four functional degrees of freedom, consistent with the TwoNN intrinsic-dimensionality estimate of $d = 4.5 \pm 0.1$ (95\% CI).

Figure~\ref{fig:fig2}D projects the OFAT sweeps onto an orthogonalized four-axis basis constructed from the dominant PCA directions (see Methods; all six pairwise projections in Supplementary Fig.~\ref{fig:sm_ofat_projections}). Each structural factor traces a distinct, well-separated trajectory, while acquisition-related sweeps cluster near the origin. Notably, the encoder generalizes label size from a categorical training signal to a continuous physical property: training sampled only three discrete sizes (150, 210, and 330\,bp), yet the encoder organizes them onto a continuous coordinate spanning the full sweep range (Fig.~\ref{fig:fig2}D). The encoder thus recovers this coordinate structure without ever being supervised on the underlying factors.

\begin{figure}
    \centering
    \includegraphics[width=1\linewidth]{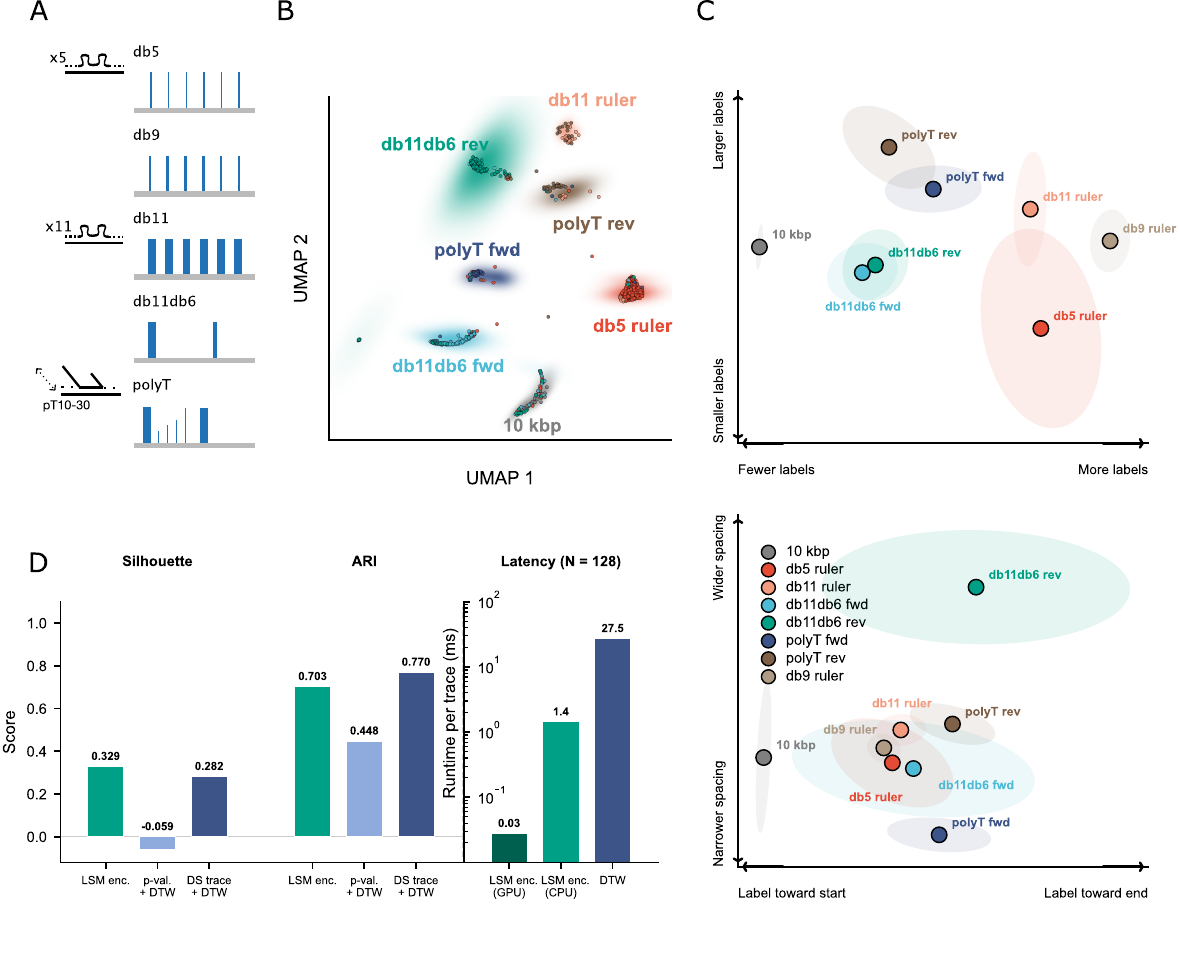}
 \caption{\small \textbf{Experimental validation of the synthetic coordinate system.}
(A) Library of dsDNA barcode molecules used in this study. The symmetric rulers db5/db9/db11 carry six dumbbell labels of increasing size, db11db6 is an asymmetric pair, and polyT carries four poly-thymine overhangs (pT10–pT30) between two db11 labels.
(B) Experimental traces projected into the encoder's latent space. UMAP projection where density contours show the embeddings of synthetic molecules whose barcode patterns match the corresponding real molecules, and overlaid points are experimental translocation traces. No retraining or fine-tuning was applied.
(C) Experimental molecule centroids projected into the functional coordinate system derived from synthetic data (Fig.~\ref{fig:fig2}D). Each centroid is plotted with a 1 std-dev covariance ellipse across all six pairwise projections of the four functional coordinate axes (Supplementary Fig.~\ref{fig:sm_ofat_molecules}).
(D) Benchmarking against dynamic time warping (DTW) baselines. Left: silhouette score. Center: adjusted Rand index (ARI). Right: runtime per trace for a bag of 128 events. LSM encoder, DTW, and p-values + DTW are compared in each panel.}
    \label{fig:fig3}
\end{figure}

\subsection{The Coordinate System Transfers to Real Molecules}

The OFAT analysis establishes the encoder's coordinate system on synthetic data; the critical test is whether that same system correctly organizes real molecules the encoder has never seen during training. To verify the latent space mapping, we performed translocation experiments of different types of engineered dsDNA structural barcodes with nanopores ranging from 7 to 28 nm in diameter (see Methods for acquisition and event curation details). Individual traces of five distinct molecules were acquired using a 10 MHz nanopore reader system\cite{Cartiglia2026}:  three symmetric rulers (DB5, DB9, DB11), each carrying six labels of identical size but differing in dumbbell dimensions; an asymmetric molecule combining labels of different sizes (DB11DB6); and a mixed-chemistry molecule combining polyT and dumbbell labels on the same scaffold (polyT)(Fig.~\ref{fig:fig3}A), alongside a bare 10kbp dsDNA control. 

Embeddings are projected to two dimensions with UMAP~\cite{mcinnes2018umap} and clustered with DBSCAN (see Methods). For non-palindromic molecules such as DB11DB6 and polyT, forward and reverse translocation orientations naturally form distinct clusters (Supplementary Section~\ref{sec:sm_clusterin}).

UMAP was fitted on encoder embeddings of synthetic traces matching the five experimental molecules; experimental traces were then projected into the same space using the frozen transform (Fig.~\ref{fig:fig3}B). Without retraining, they cluster in the positions predicted by the synthetically trained encoder. In the OFAT-derived coordinate system (Fig.~\ref{fig:fig3}C), the three rulers co-localize at the high label-count end but separate monotonically along the label-size axis (DB5\,<\,DB9\,<\,DB11), demonstrating that the encoder resolves label number and label size independently. Asymmetric molecules separate along the label-position axis according to translocation orientation, and the featureless 10kbp maps to the periphery. That a coordinate system learned entirely from synthetic data correctly stratifies real molecules confirms the encoder has captured genuine physical structure.

We compared the LSM encoder to two DTW baselines (see Methods) on experimental data pooled across different chips (Fig.~\ref{fig:fig3}D). Both baselines use the same DTW algorithm but with different inputs: the raw downsampled trace and the p-value representation (Methods). The encoder produces the most compact, well-separated clusters, achieving the highest silhouette score (0.329). Its adjusted Rand index (ARI) is 0.703, within reach of downsampled-trace DTW (0.770); the small gap reflects DBSCAN boundary placement rather than a difference in underlying cluster geometry (Supplementary Table~\ref{tab:clustering_results}). Unlike the encoder, DTW has no transferable representation: it requires labeled reference traces, is sensitive to acquisition conditions, and cannot pool data across devices. The LSM encoder is also ${\sim}20\times$ faster on CPU and ${\sim}1000\times$ faster on GPU (Fig.~\ref{fig:fig3}D \& Supplementary Fig.~\ref{fig:sm_latency}). Critically, this performance is achieved without any per-experiment labeling or reference traces: events are embedded directly as acquired, enabling real-time classification on a multichannel nanopore reader system~\cite{Cartiglia2026} (Methods; Supplementary Video~1).

\begin{figure}
\centering
\includegraphics[width=0.95\linewidth]{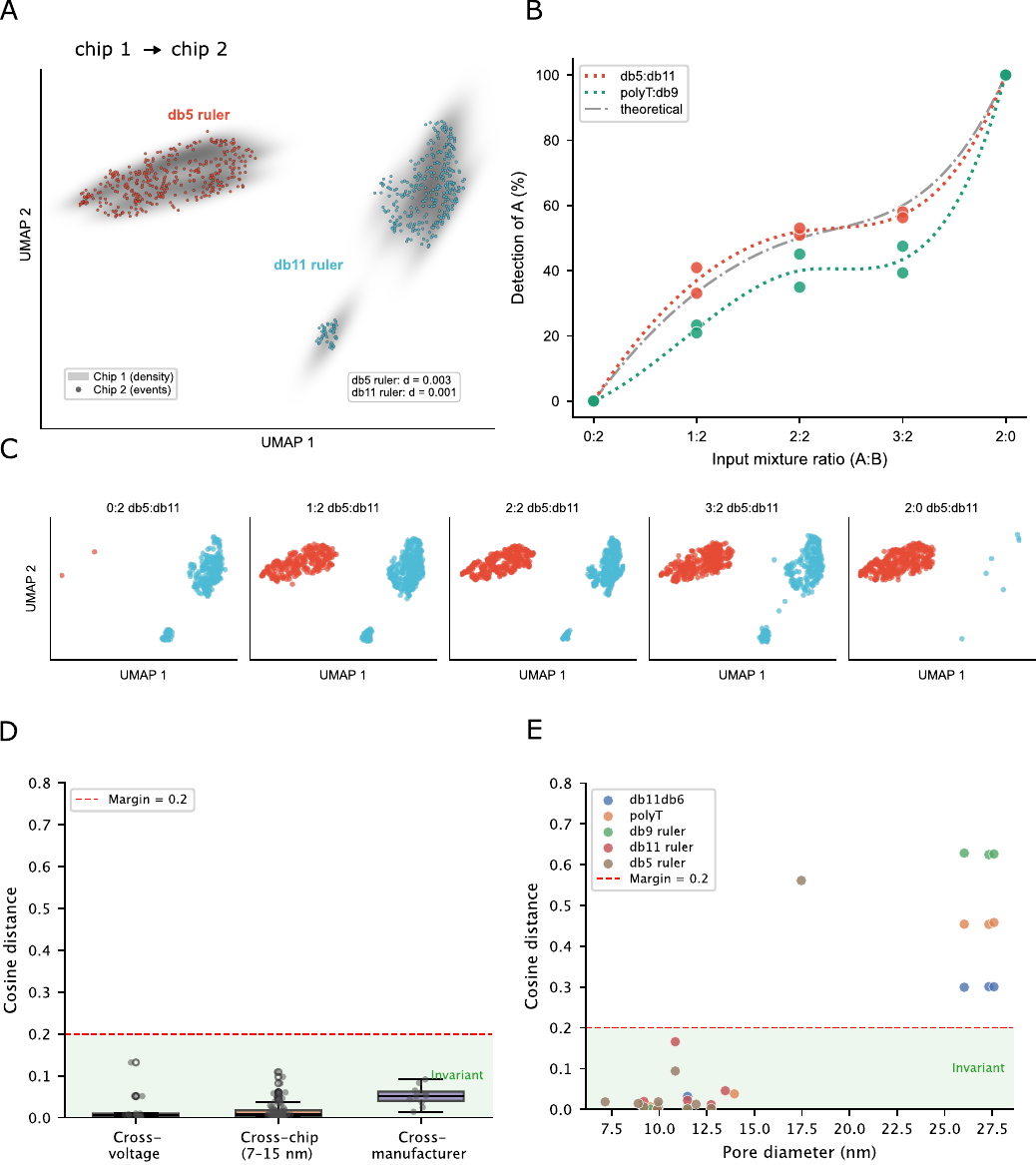}
\caption{\small \textbf{Encoder invariance across acquisition conditions and quantitative mixture analysis.}
(A) Cross-device projection. Chip~2 events are projected into Chip~1's UMAP space without retraining. Cluster centroid shifts between the two chips are d=0.003 (DB5) and d=0.001 (DB11) confirm near-perfect cross-device alignment.
(B) Mixture quantification for two binary mixtures (DB5:DB11 and polyT:DB9) at five volumetric ratios on two distinct pores.
(C) UMAP projections of the DB5:DB11 mixture at each ratio.
(D) Embedding stability across voltage, chip, and manufacturer. Each point shows the cosine distance between a per-condition centroid and a same-molecule reference. The analysis spans 19 independent chips from two manufacturers and two fabrication methods, five molecule types, and three voltages (over 100 unique experimental conditions). The dashed line marks the contrastive margin of $m=0.2$.
(E) Embedding stability as a function of pore diameter. Each point shows the cosine distance between a molecule's embedding on a given chip and a same-molecule reference measured on a different chip. The dashed line marks the contrastive margin of $m=0.2$.}
\label{fig:fig4}
\end{figure}

\subsection{Device Invariance Enables Cross-Experiment Pooling}
\label{sec:mixtures}

A molecular coordinate system is only useful if it remains stable when acquisition conditions change. In nanopore sensing, voltage, device, fabrication process, and pore diameter all vary between experiments. If the representation shifts with these conditions, each chip requires independent calibration.

As a first demonstration of acquisition invariance, we transferred the representation between two independent chips. We fitted the UMAP manifold on data from a reference chip (Chip~1) with frozen encoder weights, then projected translocation events from an independent Chip~2 directly into this space (Fig.~\ref{fig:fig4}A). Cluster centroids shift by only $d = 0.003$ (DB5) and $d = 0.001$ (DB11) in cosine distance, confirming near-perfect cross-device alignment. To our knowledge, cross-device pooling of this kind has not previously been demonstrated for solid-state nanopore single-molecule identification.

Because embeddings transfer across chips, events from different devices can be pooled and analyzed together, enabling quantitative mixture analysis. Two binary mixtures (DB5:DB11 and polyT:DB9), each at five volumetric ratios on two chips, were quantified by assigning traces to their nearest cluster centroid (Fig.~\ref{fig:fig4}B,C). Detected proportions closely track input ratios for DB5:DB11; the polyT:DB9 pair shows a systematic offset at intermediate ratios, possibly reflecting differences in capture probability or detection sensitivity between the two molecules. Notably, DB5 and DB11 differ only in the number of dumbbells, confirming the fine resolution of the coordinate system (Supplementary Fig.~\ref{fig:sm_polyT_ratios} and Section~\ref{sm:quantification}).

The two-chip transfer and mixture results above hold only if invariance generalizes beyond a single pair of devices. We therefore tested it systematically across 19 chips from two manufacturers, five molecule types, and three voltages, comprising over 100 unique conditions (Fig.~\ref{fig:fig4}D, Supplementary Table~\ref{tab:invariance_datasets}). Changing voltage (0.2--0.4\,V), swapping chips (7--15\,nm), or switching manufacturer (NORCADA vs.\ imec) all produce centroid shifts below the contrastive margin (Supplementary Fig.~\ref{fig:sm_pore_invariance}).
The pore diameter dependence reveals a physical boundary of this invariance (Fig.~\ref{fig:fig4}E). Embeddings remain stable for pores in the 7--15\,nm range regardless of molecule identity, but diverge at 25\,nm, where signal amplitudes fall outside the synthetic training distribution ($\Delta I/I_0 \in [0.03, 0.08]$, corresponding to roughly 8--13\,nm pore diameters).

\begin{figure}
\centering
\includegraphics[width=\linewidth]{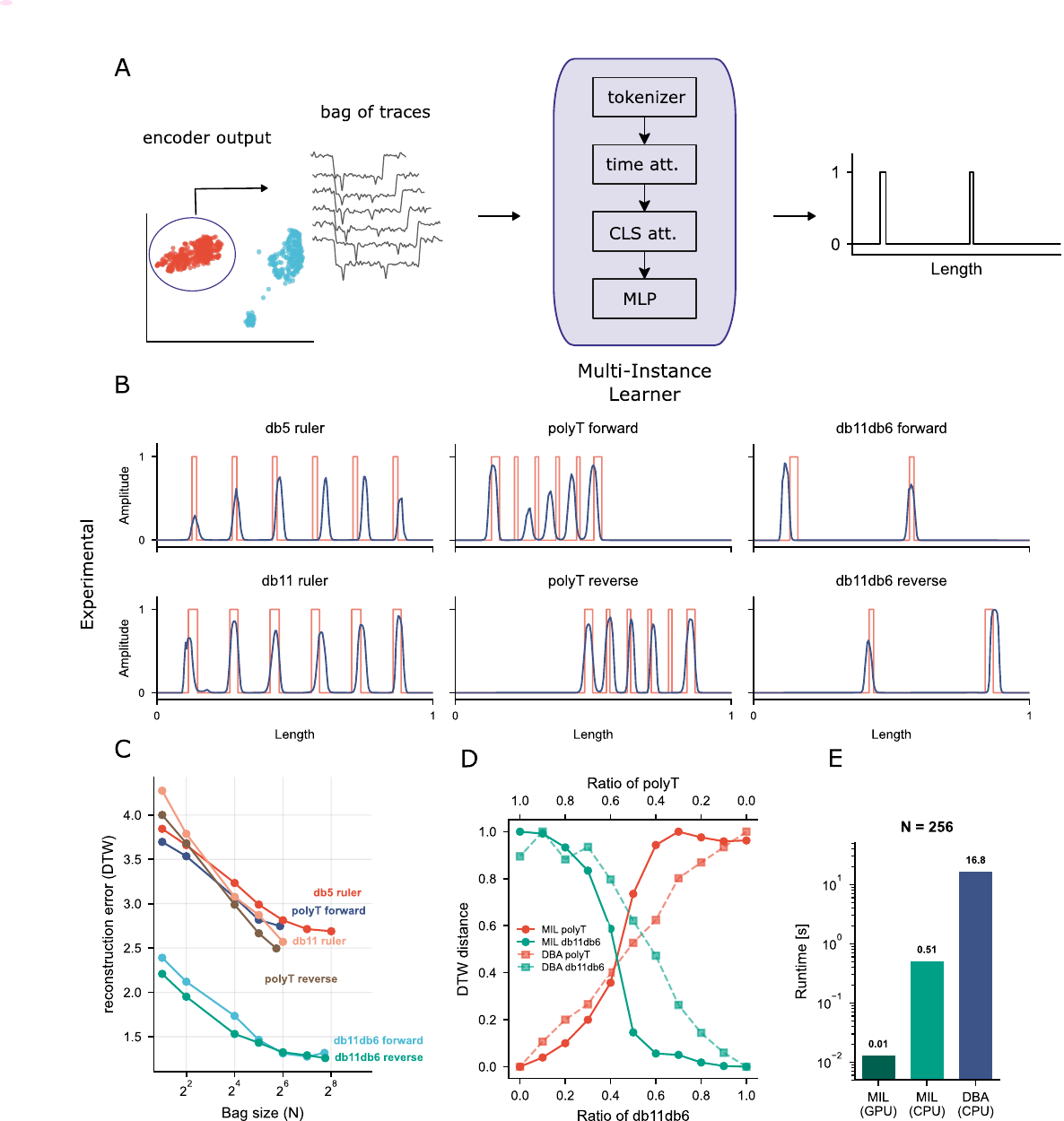}
\caption{\small \textbf{Multi-Instance Learner consensus reconstruction.}
(A) Schematic of the MIL pipeline. Traces from an encoder-identified cluster are converted to a noise-normalized p-value representation and processed by a Transformer-based Multi-Instance Learner, which alternates between trace-level refinement (Time Attention) and bag-level aggregation (CLS Attention) to output a single consensus binary barcode.
(B) Reconstruction results for six experimental molecule classes. Each panel shows the ground-truth barcode (red) alongside the MIL consensus prediction (blue) for DB5, DB11, DB11DB6, and polyT (the latter two in both forward and reverse translocation orientations).
(C) Reconstruction fidelity as a function of bag size. Dynamic time warping (DTW) distance between the MIL reconstruction and the ground truth, evaluated across all six experimental clusters (four molecules, two giving separate forward and reverse orientation clusters).
(D) Robustness to cluster contamination. DTW distance of MIL and DTW barycenter averaging (DBA) reconstructions as a function of the contamination ratio between DB11DB6 and polyT traces in the input bag.
(E) Runtime comparison using
N=256 traces. MIL runtime on CPU and GPU compared against DBA on CPU.}
\label{fig:fig5}
\end{figure}

\subsection{Reconstructing the Digital Barcode}

Clustering identifies which traces share a molecular identity, but it does not recover the barcode itself. To reconstruct the underlying barcode from a cluster of stochastically warped events, our LSM approach is extended with a Transformer-based Multi-Instance Learner (MIL). 
Traces from a single encoder-identified cluster are first converted to a noise-normalized p-value representation, a compact tokenization that scores each of 128 temporal windows for the presence of a label (see Methods). They are then processed by the MIL, which alternates between trace-level refinement (Time Attention) and bag-level aggregation (CLS Attention) to output a single consensus binary barcode (Fig.~\ref{fig:fig5}A). 

The encoder's unsupervised clustering is a necessary prerequisite, ensuring that each bag predominantly contains traces of one molecular identity, with a fraction of contaminating traces tolerated through MIL training (see Methods).
The MIL recovers accurate consensus barcodes across the symmetric rulers (DB5, DB11), the asymmetric DB11DB6, and the mixed-chemistry polyT, in both translocation orientations (Fig.~\ref{fig:fig5}B). Applied independently to each encoder-identified cluster within the binary mixture experiments, the MIL correctly recovers the barcode of every component species (Supplementary Fig.~\ref{fig:sm_mixture_reconstructions}). Reconstruction fidelity improves consistently as more traces are added to the bag (Fig.~\ref{fig:fig5}C), confirming that the MIL pools information across events to sharpen the consensus. 

We benchmarked the MIL against DTW barycenter averaging (DBA)~\cite{petitjean2011global}, a time-domain consensus method based on iterative DTW alignment (see Methods). Unlike DBA, whose reconstruction quality plateaus with bag size (Supplementary Fig.~\ref{fig:sm_dba_scaling}), the MIL continues to improve as more traces are added, and it degrades more gracefully under cluster contamination (Fig.~\ref{fig:fig5}D). It is also ${\sim}33\times$ faster on CPU and ${\sim}1300\times$ faster on GPU (Fig.~\ref{fig:fig5}E; Supplementary Section~\ref{sec:sm_mil_latency}).

\begin{figure}
\centering
\includegraphics[width=\linewidth]{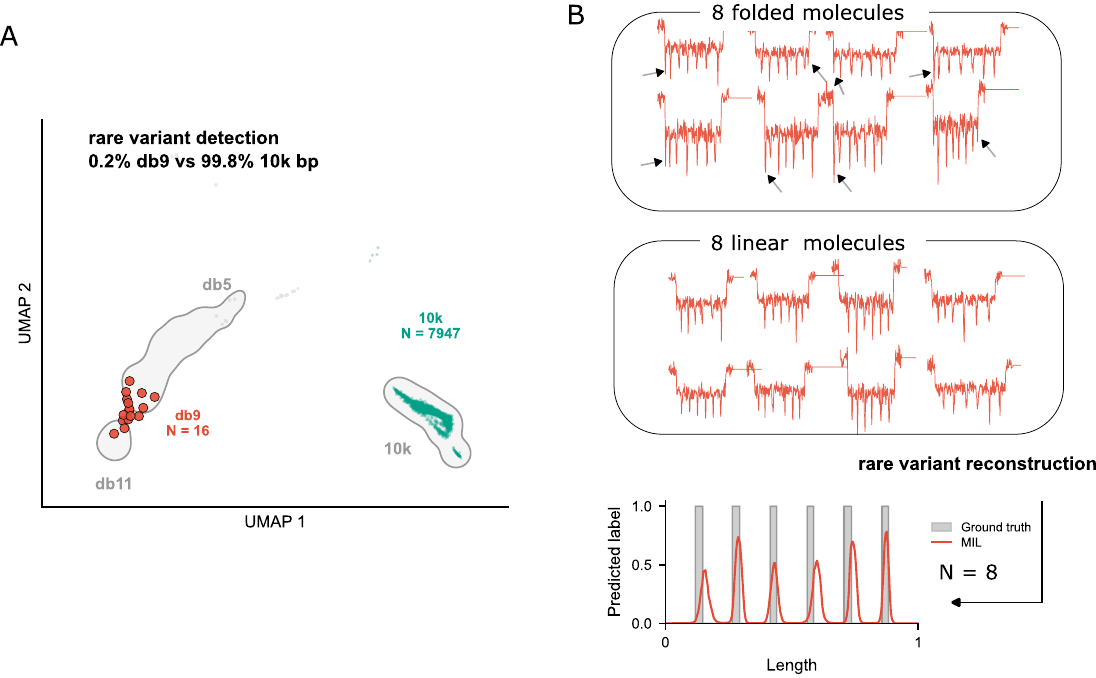}
    \caption{\small \textbf{Detection and identification of a rare molecular variant at 0.2\% abundance.} 
    (A) Reference UMAP projection fitted on pooled calibration experiments of DB5, DB11, DB11DB6, polyT, and 10,kbp unlabeled dsDNA, with DB9 excluded from the fit; density contours are shown for DB5, DB11, and 10,kbp. Translocation events from the rare-variant mixture ($\sim$0.2\% rare variant, four independent pores) are projected into this UMAP. Majority-population events are shown in green; 16 rare-variant events are highlighted in red. 
    (B) Individual translocation traces of the 16 detected rare-variant events, comprising 8 folded and 8 linear translocations; arrows mark the fold-induced current. The MIL consensus barcode reconstructed from the 8 linear events is shown at the bottom.}
      \label{fig:fig6}
\end{figure}

\subsection{Detection of Rare Variants}\label{sec:rare_variants}

The structured geometry of the learned coordinate system makes rare variant detection a natural consequence: a molecule with structure occupies a distinct region of the latent space simply because its physical degrees of freedom differ from those of the background population.
We tested this using DB9. Like the other rulers, it carries six labels, but at a distinct dumbbell size (9 repeats) that was excluded from both the encoder's synthetic training set and the UMAP reference frame, which was fitted only on DB5, DB11, DB11DB6, polyT, and unlabeled 10,kbp dsDNA (only DB5, DB11, and 10,kbp are shown in Fig.~\ref{fig:fig6}A for clarity). When DB9 events are projected into this pre-fitted frame, they fall between the DB5 and DB11 clusters (Fig.~\ref{fig:fig6}A).

Detecting a rare variant against an overwhelming background is the regime where geometric, single-event detection should pay off: because identity is read from position in the latent space rather than from cluster statistics, a single correctly placed event is in principle sufficient. To test this limit, we spiked DB9 into unlabeled 10,kbp DNA at 0.2\% molar abundance, roughly one target molecule in 500, and measured the mixture on four independent pores in parallel, pooling the events into the frozen UMAP reference frame to leverage the cross-experiment invariance established above. 
The 10,kbp majority population remained within its pre-defined boundary, while 16 events fell entirely outside it, localizing precisely in the expected DB9 zone (Fig.~\ref{fig:fig6}A). 
To confirm these 16 events are not chance placements of background molecules, we projected an independent negative control of 3,461 unlabeled 10\,kbp events into the same frame: none fell in the DB9 region, bounding the per-event false-positive rate below $10^{-3}$ (see Methods and Supplementary Section~\ref{sec:sm_fpr}).

These 16 events comprised 8 linear and 8 partially folded translocations (Fig.~\ref{fig:fig6}B). Both conformations co-localized within the same latent sub-region, and all 16 traces are visually consistent with the DB9 barcode structure. Together with the background-only negative control, these independent lines of evidence converge on a single DB9 identity. With the variant identified, the MIL reconstructs its consensus barcode directly from the 8 linear events (Fig.~\ref{fig:fig6}, bottom), recovering the full label pattern from a population this rare.

Beyond the 16 DB9 detections, 23 further events localized away from both the 10,kbp background and the DB9 region, and their current signatures are structurally inconsistent with DB9 (Supplementary Section~\ref{sec:sm_outliers} and Fig.~\ref{fig:sm_outliers}); they are most plausibly trace-level contamination from co-processed analytes or atypical translocations. In every case, they fall outside the DB9 detection boundary, leaving the rare-variant count unaffected.

\section{Conclusion}

In summary, we show that stochastic solid-state nanopore traces can be converted into an interpretable molecular coordinate system by training a contrastive encoder on physics-informed synthetic data. In this representation, individual translocation events are no longer analyzed as variable time-domain waveforms but as points in a learned space where detection is geometric and identity is determined by position. Under the conditions tested here, this space is stable across nanopore sensors, allowing events from different pores to be pooled directly without temporal averaging or pore-specific retraining. More generally, any high-bandwidth and stochastic pulse-based sensing method with a tractable model could, in principle, be represented in the same way, e.g. biological nanopore squiggles, single-molecule fluorescence, force spectroscopy traces, or single-entity electrochemical measurements.

We introduce a paradigm shift where the learned space reflects the physical degrees of freedom of the analyte and its translocation dynamics. This distinguishes Latent Space Mapping (LSM) from nanopore machine-learning approaches focused on supervised classification~\cite{misiunas2018quipunet, bao2021squigglenet, cao2024deep}, feature extraction~\cite{dematties2021deep}, protein fingerprinting~\cite{soni2025protein}, or physics-guided interpretation of input features~\cite{shah2026interpretable}.  In contrast, LSM makes interpretability a property of the learned coordinate system itself, not of the classifier output or engineered features.  

This representation is also operationally useful. 
A single encoder pass is sufficient for inference, making the framework compatible with real-time analysis of high-throughput nanopore arrays. The synthetic training strategy further enables candidate barcode designs to be evaluated in silico before synthesis by projecting their expected signals into latent space and estimating their separability. Although the present forward model describes translocation stochasticity through empirically sampled velocity profiles, richer Langevin-based or molecular dynamics models should extend the approach to broader molecular regimes. These results establish physics-informed contrastive learning as a route to scalable, molecularly interpretable, single-event sensing with solid-state nanopores.

\section{Methods}

\subsection{Design and Synthesis of Barcoded DNA Nanostructures}

All analytes are built on a common 7,228\,bp double-stranded DNA scaffold purchased from Cambridge Nucleomics. Two label chemistries are employed: dumbbell (DB) labels, in which a short duplex hairpin is covalently attached to the scaffold backbone, and polyT (pT) labels, in which a single-stranded thymine overhang is appended. Five distinct barcoded molecules were studied. The ruler series: DB5, DB9, and DB11, each carry six DB labels of the corresponding size distributed along the scaffold, providing a symmetric, periodically patterned barcode. The molecule DB11DB6 carries two labels of different sizes (one DB11 and one DB6), creating an asymmetric, non-palindromic barcode; both translocation orientations are observed experimentally and treated as distinct events. The molecule polyT carries four pT labels of increasing length (pT10, pT15, pT20, pT30) flanked by two DB11 labels, forming a six-label asymmetric barcode. A bare 10\,kbp linear dsDNA strand with no attached labels serves as an unlabeled control.

\subsection{Solid-State Nanopore Fabrication and Data Acquisition}\label{sec:methods_acquisition}

Measurements were performed on solid-state nanopores fabricated in silicon nitride ($\text{SiN}_x$) membranes from two sources: commercial membranes from Norcada (nominal pore diameters of \SI{10}{nm} and \SI{25}{nm}) and EUV-lithography-fabricated membranes from imec~\cite{chaudhuri2025fabrication}. In total, the experimental dataset spans 10 Norcada chips with pore diameters in the 7--15\,nm range, 3 Norcada chips with \SI{25}{nm} nominal pores, and 6 imec chips, across five molecule types (Supplementary Table~\ref{tab:invariance_datasets}). The nanopore chips were mounted in a custom flow cell, and both the \textit{cis} and \textit{trans} chambers were filled with a measurement buffer consisting of \SI{4}{M} LiCl.

Bias voltages of \SI{0.2}{V}, \SI{0.3}{V}, and \SI{0.4}{V} were applied across the pore and the ionic current was read using Ag/AgCl electrodes. The resulting signal was recorded using a four-channel Elements s.r.l. acquisition system, with each channel operating at a high-bandwidth sampling rate of \SI{10}{MHz}.
Event detection of the raw translocation traces was performed using the GPU-accelerated Data Sieving framework~\cite{Cartiglia2026}. Following event detection, a minimum area-under-the-curve (AUC) threshold of \SI{0.2}{pC} in the dwell-time--current-depth plane was applied to reject transient noise spikes. The blockage-level distribution of retained events was then fitted with a kernel density estimate and a Gaussian mixture model to assign each event a morphological label (linear, proximally folded, fully folded, knotted, etc), verified by visual inspection of representative traces. Only linear events were retained for the main analyses.

\subsection{Synthetic Dataset Generation and Data Augmentation}\label{sec:methods_synthetic}

Both models were trained exclusively on physics-informed synthetic data, with no experimental traces used during training. The synthetic data generation proceeds in three stages. First, a library of 4,000 unique reference barcodes is generated algorithmically, each represented as a binary vector at 30\,bp resolution over the 7,228\,bp molecular length. Second, for each barcode, stochastic translocation traces are simulated: a non-linear velocity profile sampled from the empirically measured distribution (rather than generated by a dynamical translocation model; rationale in Supplementary Section~\ref{SubSub:VelocityProfiles}) stretches and compresses the barcode template from base-pair space into the time domain, with a fraction of traces generated in a folded conformation. Third, the time-domain signal is convolved with a pore-specific impulse response and combined with noise matched to the measured power spectral density, with the signal depth $\Delta I/I_0$, signal-to-noise ratio, and label amplitude sampled independently per trace.

Linear and folded translocation events are both generated, so that both conformations are represented during training. All sampled parameters and their distributions are calibrated against experimental DNA ruler measurements and summarized in Supplementary Table~\ref{tab:synth_params}; full pipeline details are provided in Supplementary Section~\ref{sec:sm_synthetic_detail}.

\subsection{Siamese Encoder}\label{sec:methods_encoder}

Each translocation trace is low-pass filtered at \SI{250}{kHz} with a fourth-order Butterworth filter, decimated at a fixed rate and zero-padded to 650 samples, and mapped into a 256-dimensional embedding by a 1D ResNet encoder~\cite{he2016deep, bao2021squigglenet}. The encoder consists of an initial convolutional layer producing 64 channels, followed by a stack of residual blocks that progressively halve the temporal dimension and double the channel count up to 1024 channels. Global average pooling over the temporal axis yields a fixed-length feature vector, which is projected to the final 256-dimensional embedding through a multilayer perceptron ($1024 \rightarrow 512 \rightarrow 256$, ReLU activations).

The encoder is trained in a Siamese configuration with a contrastive cosine similarity loss:
\begin{equation}
\mathcal{L}(x_1, x_2, y) =
\begin{cases}
1 - \cos(x_1, x_2), & y = 1 \\[6pt]
\max\left(0, \cos(x_1, x_2) - m \right), & y = -1
\end{cases}
\end{equation}
where $x_1$ and $x_2$ denote the embeddings of two traces, $y \in \{1,-1\}$ indicates whether the traces share the same ground-truth barcode, and $m = 0.2$ is the separation margin. The loss is computed over all pairs within each batch and averaged. The model is trained on both linear and folded synthetic events, so that folded translocations map to the same latent region as their linear counterparts rather than being discarded.

Training was performed for 200 epochs with a batch size of 800, using the Adam optimizer with a learning rate of $10^{-3}$ and a dropout rate of 0.2. During training, a Brownian-motion time-warping augmentation~\cite{kipen2025brownian} resamples each trace along a stochastically warped time axis, exposing the encoder to a broader range of velocity fluctuations than the synthetic profiles alone provide and strengthening its velocity invariance. The dataset was split by barcode identity (85\% training, 15\% validation), ensuring that no ground-truth barcode appears in both sets. Validation curves and full architectural specifications are provided in Supplementary Section~\ref{sec:sm_encoder}.

\subsection{One-Factor-at-a-Time (OFAT) Analysis}\label{sec:methods_ofat}

To interrogate the internal structure of the learned latent space, we performed one-factor-at-a-time (OFAT) sweeps through the synthetic parameter space. Ten generative factors were analyzed: five barcode identity factors (label count, label size, single label position, skewness, and inter-label distance) and five acquisition nuisance factors (fold fraction, signal depth, signal-to-noise ratio, translocation speed, and label amplitude). For each sweep, the factor under study was varied systematically across its full range while the remaining factors were sampled from their default distributions. At each level, 50 synthetic traces were generated and passed through the frozen encoder, and their embeddings were averaged into a single per-level mean embedding.

Each factor's sweep traces a path through the 256-dimensional embedding space. To extract the dominant direction of this path, the per-level mean embeddings were centered and the first principal component computed, yielding a unit vector that captures the direction of maximum displacement as that factor changes. The sign of each axis was chosen so that increasing factor values correspond to forward displacement along it. Encoder responsiveness (Fig.~\ref{fig:fig2}B) was quantified as the maximum pairwise cosine distance between any two per-level means along each sweep. We adopt the training margin (m=0.2) as a natural reference scale: a maximum cosine distance above 0.2 indicates that the encoder discriminates between levels of that factor, while a value near zero indicates invariance.

To assess whether the five structural axes are geometrically independent, we computed their pairwise cosine similarities (Fig.~\ref{fig:fig2}C). Values near zero indicate that the encoder represents those factors in orthogonal subspaces, while values approaching one indicate shared encoding directions. Most of the ten pairwise comparisons are nearly orthogonal ($\leq 0.21$). The exception is the pair of skewness and inter-label distance (cosine similarity 0.72), which is expected: both factors describe the relative spatial arrangement of labels on the scaffold and therefore share geometric content.

To construct a four-dimensional functional basis for visualization (Fig.~\ref{fig:fig2}D), the skewness and inter-label distance axes were averaged into a single geometry direction. The resulting four directions, label count, label size, label position, and label geometry, were orthogonalized via the Gram-Schmidt procedure. The basis is insensitive to axis ordering because label count, label size, and label position are nearly orthogonal to all other factors. Each OFAT sweep in Fig.~\ref{fig:fig2}D is projected onto pairs of these orthogonalized axes, producing the two-dimensional slices of the latent coordinate system. 

The OFAT analysis measures each factor's effect on the embedding one at a time, building an intuitive picture of how individual factors are represented. However, it cannot capture couplings between factors that arise only when multiple factors vary simultaneously. The intrinsic-dimensionality estimate complements OFAT by probing the embedding as a whole. Intrinsic dimensionality was estimated independently using the TwoNN estimator~\cite{facco2017estimating} on unit-sphere-projected embeddings, with bootstrap confidence intervals computed over 1,000 resamples. Full sweep parameters, basis construction details, and dimensionality estimation are provided in Supplementary Section~\ref{sec:sm_encoder}.

\subsection{Dimensionality Reduction and Clustering}\label{sec:methods_clustering}

Latent space clusters are identified by projecting the 256-dimensional embeddings into two dimensions with UMAP using cosine distance as the metric and applying density-based clustering with DBSCAN. UMAP is a nonlinear dimensionality reduction method that preserves local neighborhood structure, so that geometrically similar embeddings remain close in the projection. DBSCAN is a density-based clustering algorithm that identifies clusters of arbitrary shape without requiring the number of clusters to be specified in advance, and naturally designates low-density regions as noise.
Dimensionality reduction is necessary because pairwise distances concentrate in high-dimensional spaces, making density-based methods unreliable. For non-palindromic molecules such as DB11DB6 and polyT, forward and reverse translocation orientations produce distinct embedding clusters and are treated as separate groups in downstream analysis. Full parameter settings are provided in Supplementary Section~\ref{sec:sm_clusterin}.

\subsection{Clustering Evaluation Metrics}

Clustering quality is assessed using two complementary metrics. The silhouette score measures how well each data point fits within its assigned cluster relative to neighboring clusters. For each point, it computes the ratio between the mean distance to points in the same cluster and the mean distance to points in the nearest competing cluster, yielding a value between $-1$ (misclassified) and $+1$ (well-clustered); a negative score indicates that clusters overlap in the embedding space. The Adjusted Rand Index (ARI) measures the agreement between predicted cluster assignments and known ground-truth labels. It counts the fraction of data point pairs that are correctly grouped together or correctly separated, with a correction for chance agreement, yielding 0 for random labeling and 1 for perfect agreement.

\subsection{Invariance Quantification}\label{sec:methods_invariance}

Each experimental condition is a unique (molecule, chip, voltage) tuple. For a given condition, all translocation events are embedded by the frozen encoder and $\ell_2$-normalized; the condition centroid is the $\ell_2$-normalized mean of these unit vectors. For each molecule, a single anchor centroid is fixed by selecting a randomly chosen chip within the 7--15\,nm diameter bracket at the lowest voltage. The cosine distance $d = 1 - \mathbf{c}_\mathrm{condition} \cdot \mathbf{c}_\mathrm{anchor}$ between every other condition centroid and the anchor is then compared against the contrastive margin ($m = 0.2$). Conditions are grouped as cross-voltage (same chip, different voltage), cross-chip (different chip, same manufacturer), or cross-manufacturer (different manufacturer). The pore diameter analysis (Fig.~\ref{fig:fig4}E) pools all voltages per chip into a single centroid to isolate the effect of pore size. Full details, including orientation assignment for asymmetric molecules, are provided in Supplementary Section~\ref{sec:sm_invariance_methods}.

\subsection{Baseline Methods: Dynamic Time Warping and Barycenter Averaging}\label{sec:methods_dtw}

We benchmark the LSM framework against two classical time-domain approaches. Dynamic Time Warping (DTW) is an alignment algorithm for temporal sequences that accounts for non-linear differences in timing. Given two traces, DTW finds the alignment that minimizes the cumulative point-wise distance, allowing one trace to be locally stretched or compressed to match the other; the resulting DTW distance quantifies how dissimilar two traces are after optimal temporal alignment. For the clustering baseline, pairwise DTW distances are computed between all traces and used as input to UMAP and DBSCAN in place of the encoder's cosine distances.

A key limitation of DTW is that it has no learned or transferable representation: the dissimilarity score is recomputed pairwise from raw traces every time, so classifying a new molecule requires labeled reference traces from that molecule. DTW distances also reflect acquisition conditions (pore geometry, applied voltage, noise characteristics) in addition to molecular structure. Consequently, the same molecule measured on two different chips can produce different DTW distances, fragmenting any attempt at cross-device data pooling.

DTW Barycenter Averaging (DBA) extends DTW from pairwise comparison to consensus estimation. It computes an average trace from a set of input sequences by iteratively refining a candidate average: in each iteration, every input trace is DTW-aligned to the current candidate, and each time point of the candidate is updated to the mean of all samples aligned to it. DBA serves as the time-domain baseline for the MIL's barcode reconstruction task.

\subsection{Multi-Instance-Learner}

The MIL receives a bag of $N$ p-value feature vectors from traces assigned to the same latent cluster and outputs a single binary consensus barcode. Each of the $T = 128$ p-values in a trace is treated as an individual token and projected into a $C = 256$-dimensional embedding. The resulting bag of $N$ token sequences is processed by a Transformer that alternates between two attention operations: Time Attention, which applies self-attention independently within each trace to refine per-position representations; and CLS Attention, which applies self-attention across all traces via learnable classification tokens to aggregate bag-level information. A global output token $\mathbf{c}_{\mathrm{out}}$ accumulates bag-level consensus across $L = 4$ layers and is decoded by a linear layer into the $B = 241$-bin barcode prediction. No positional encoding is applied across traces, making the architecture permutation-invariant with respect to trace ordering~\cite{zaheer2017deep, lee2019set}.

The Transformer uses $H = 4$ attention heads and a feedforward hidden dimension of $512$. The model is trained for 200 epochs with a batch size of 800, using the Adam optimizer with a learning rate of $10^{-4}$, a dropout rate of 0.2, and binary cross-entropy loss. Each training bag contains 32, 64, or 128 traces sampled from the same ground-truth barcode, with up to 25\% of traces replaced by traces from a different ground truth to improve robustness against imperfect clustering. The dataset was split by barcode identity (70\% training, 30\% validation), ensuring no overlap between sets. Full architecture specifications and validation curves are provided in Supplementary Section~\ref{sec:sm_MIL}.

\subsection{p-value Feature Extraction}\label{sec:methods_pvalue}

Instead of operating on raw current traces, the MIL takes as input a compact, noise-normalised representation of each event. The event interval is partitioned into 128 equal-length temporal windows, and within each window, the statistical significance of the observed current minimum is evaluated against a trace-specific Gaussian noise model estimated from the median and median absolute deviation of the event signal. The resulting p-value quantifies how unlikely the observed current depression is under the noise-only null hypothesis, and is corrected for the reduced number of independent samples introduced by low-pass filtering. Small p-values therefore mark candidate label positions, providing a representation that is invariant to baseline current level and RMS noise. The full mathematical derivation is given in Supplementary Section~\ref{sec:P-Value}.

\subsection{Real-Time Preprocessing}\label{sec:methods_realtime}

For real-time operation, linear events are identified on-the-fly from the incoming current--dwell-time scatter plot using the population statistics. Each detected event is trimmed~\cite{Cartiglia2026} to its blockage interval and retained unfiltered. The raw trace is then low-pass filtered at \SI{250}{kHz} with a finite impulse response (FIR) filter, downsampled to 650 samples, and passed to a compiled (DLL) implementation of the encoder for immediate embedding. Real-time operation is demonstrated in Supplementary Videos.

\subsection{Rare-Variant False-Positive Rate}\label{sec:methods_fpr}
 
To estimate how often background events might be mistaken for rare-variant detections, we measured a negative-control sample of unlabeled 10\,kbp dsDNA containing no DB9: $n = 3{,}461$ events across 12 independent chips (6 Norcada, 6 imec; 0.2--0.4\,V). Each event was embedded and projected into the rare-variant UMAP frame. The DB9 detection region was fixed from the calibration frame ($\mathrm{UMAP}_1 < -2.5$, $\mathrm{UMAP}_2 < 0$). None of the 3,461 background events landed in this region. Because we observed zero false positives, the standard Poisson ``rule of three'' gives a 95\% upper bound on the expected count of $\lambda_{95} = -\ln(0.05) \approx 3.0$, so the per-event false-positive rate is at most $p_\mathrm{FP} < 3.0/3{,}461 \approx 8.7 \times 10^{-4}$. Applied to the rare-variant experiment itself ($N_\mathrm{bg} = 7{,}947$ background events), this worst-case rate predicts roughly 6.9 false positives. Under a Poisson distribution with that mean, seeing 16 or more events has probability $P(X \geq 16) < 0.01$, so the 16 detections cannot plausibly be explained by background leakage into the detection zone. Full details are provided in Supplementary Section~\ref{sec:sm_fpr}.

\begin{acknowledgement}

The authors thank Silvia Lenci, Ananth Subramanian, Charlotte D’Hulst, and Paru Deshpande.

\end{acknowledgement}

\paragraph{Code Availability}
The code developed in this work is available upon reasonable request.

\paragraph{Author Contributions}
M.C., S.K., and W.B.\ contributed equally as co-first authors. 
S.M., M.C., W.B., and S.K.\ conceived the study. 
M.C.\ wrote the manuscript with input from S.M., W.B. and S.K.
M.C., W.B., and S.K.\ jointly developed and validated the computational framework, including the contrastive encoder, OFAT analysis, Multi-Instance Learner, benchmarking and rare-variant analysis, with S.K.\ implementing and training the encoder and Multi-Instance Learner.
K.O.\ and N.B.\ supported the experimental and computational infrastructure.
W.P., W.R., E.B., and L.V.\, acquired the experimental nanopore data.
P.V.D.\ provided resources and institutional support. 
S.M.\  supervised the project and provided overall project guidance. 
All authors reviewed and approved the final manuscript.

\paragraph{Declaration of competing interest}
The authors declare a competing interest: a patent application covering aspects of the work reported in this manuscript has been submitted. No other competing interests are declared.

\bibliography{achemso-demo}

\end{document}